\documentclass[aps,prl,amsmath,amssymb,reprint,superscriptaddress]{revtex4-1}
\usepackage{graphicx}
\usepackage{amsmath}
\usepackage{dcolumn}
\usepackage{bm}
\usepackage{bbold}
\usepackage{color}
\usepackage{amsfonts}
\usepackage{amssymb}
\usepackage{mathrsfs}
\usepackage{tabularx}
\usepackage{braket}
\usepackage{mathtools}
\usepackage{soul}			% For highlighting text, use command \hl
\usepackage{lettrine}

\usepackage{caption}
\usepackage{subcaption}

\begin{document} 
\title{Collective Behavior of Cr$^{3+}$ ions in Ruby Revealed by Whispering Gallery Modes}

\author{Jeremy Bourhill}
\affiliation{ARC Centre of Excellence for Engineered Quantum Systems, School of Physics, University of Western Australia, 35 Stirling Highway, Crawley WA 6009, Australia}

\author{Maxim Goryachev}
\affiliation{ARC Centre of Excellence for Engineered Quantum Systems, School of Physics, University of Western Australia, 35 Stirling Highway, Crawley WA 6009, Australia}

\author{Warrick G. Farr}
\affiliation{ARC Centre of Excellence for Engineered Quantum Systems, School of Physics, University of Western Australia, 35 Stirling Highway, Crawley WA 6009, Australia}

\author{Michael E. Tobar}
\email{michael.tobar@uwa.edu.au}
\affiliation{ARC Centre of Excellence for Engineered Quantum Systems, School of Physics, University of Western Australia, 35 Stirling Highway, Crawley WA 6009, Australia}

\date{\today}

%%%%%%%%%%%%%%%%%%%%%%%%%%%%%%%%%%%%%%%%%%%%%%%%%%%%

\begin{abstract}

\noindent \textit{We present evidence for collective action of Cr$^{3+}$ ion impurities in a highly concentrated ruby crystal coupled to microwave Whispering Gallery Modes (WGMs). 
 The cylindrical geometry of the crystal allows for the creation of superradiant, or {``}spin-mode{''} doublets, with spatial structure similar to that of WGMs. The formation of these spin patterns allows us to observe directly different selection rules namely wavenumber and azimuthal phase matching. The demonstration is made via an avoided level crossing between spin and photon mode doublets as well as absence of coupling between spin modes of different wavenumbers. The effect is observable due to strong spin-photon coupling (67 MHz) exceeding both spin ensemble and cavity losses as well as the photon doublet splitting.
  We demonstrate that a four harmonic oscillator model not only with coupling between photon resonances (0.43 MHz) but also with spin doublet (73 MHz) is necessary to accurately describe these results.}
\end{abstract}

\maketitle

%\section{Introduction}

%\section{System Description}

%\subsection{Spin-Wave interaction in highly doped crystal resonators}
%\lettrine[nindent=0em,lines=3]{I} \\

\lettrine[nindent=0em,lines=3]{S}\\
\\uperradiance is an important phenomenon in quantum optics as often the sample under study features separation distances small compared to the wavelength of exciting radiation, $\lambda$. There is renewed interest in these systems for quantum information sciences and to attain new insights into QED and its applications \cite{svidzinsky,Haroche}. For example, superradiant effects will need to be considered when constructing an optical{--}microwave interface \cite{optmic5,optmic1,optmic4} or a quantum memory \cite{optmic2,optmic3} using spin ensembles. 

Superradiance was initially defined in 1954 by Robert Dicke as the cooperative, spontaneous emission of photons from a collection of atoms \cite{dicke}. Superradiance was first observed experimentally in 1973 in the optical regime in HF Gas \cite{HF}. It has since been observed in other ultacold atomic gases \cite{He,Rb,Rb2,Na}, organic semiconductors \cite{Hagg,tetracene}, polymer thin films \cite{poly}, numerous crystalline systems \cite{YAG,crystal,crystal2}, and in artificial atoms \cite{dot1,dot2,dot3}. Here, we report the observation of superradiance in the microwave regime in a highly doped ruby sample, with relatively high concentrations of Cr$^{3+}$ ions replacing Al$^{3+}$ ions in the crystal lattice. 

In free space, when $N$ atoms are close together compared with $\lambda$, they act like one big atom and decay collectively, in phase with one another. As a result, the atoms radiate their energy $N$ times faster than for incoherent emission. A direct result is the inherent directionality associated with the emitted radiation; the emitted photons travel in the same direction as the exciting photons. This directionality is a result of the timing of the excitations; the atoms at the {``}front{''} of the sample are excited first, and those at the back, last, leading to the excitations appearing as spatial phase factors \cite{Scully}. Superradiance is a consequence of extra coherence in the system, which can be observed in additional ways on top of an increased emission rate. 

To observe coherent effects originating from collective action in the microwave regime is sufficiently more challenging than the optical regime. This is due to the relatively weak strength of field-matter interactions via magnetic fields as compared to electric fields \cite{tobar1}. When the emitters couple to a resonant cavity mode, their separation becomes irrelevant. The coherence between separate spins is generated by their interaction with a common mode, which occupies space over the entire cavity volume. The strength of this interaction is determined by the light{--}matter coupling constant, $g$, which is proportional to the concentration of the emitters \cite{ruby}.\\

%Coherence amongst emitters can only be observed if the field coupling to them is doing so with sufficient strength. This coupling is determined by, amongst other parameters, the concentration of the emitters \cite{ruby}.\\

%\noindent It is a well known phenomenon that resonant photonic whispering gallery modes (WGMs) in a circular cavity are the result of a linear superposition of clockwise and counter-clockwise circularly polarised waves. Imperfections within a cavity can result in backscattering effects that cause the degeneracy between the two waves to be lifted, by introducing a coupling between them. 

\noindent Unlike Fabry-P{\'e}rot cavities, ideal whispering gallery mode (WGM) resonators have rotational symmetry. This fact dictates that if a mode field distribution has solutions of the system eigenvalue problem, any of its rotations around the cylinder axis will also be a solution. Each of these solutions could be represented as a linear combination of only two orthogonal solutions. In actual WGM cavities, this symmetry is lifted by a number of imperfections that we further collectively call back-scatterers. These back-scatterers introduce a coupling between the two particular orthogonal solutions that depend on the back-scatterer details. A WGM will therefore manifest as two orthogonal modes (or doublets) with a difference of Sine and Cosine in the mode{'}s azimuthal dependence in its analytical expression (i.e. a difference of $\pi/2$ in azimuthal phase) \cite{backscatter}, henceforth referred to as the {``}$s${''} and {``}$c${''} modes. %Despite the orthogonality, the coupling produced by these imperfections in the crystal symmetry results in the mode appearing as a doublet. 
This manifests as a splitting of a single resonant peak into two resonant peaks by a distance equal to two times the coupling value, $\kappa$. In sapphire crystals, the losses of such WGMs are so low, that the bandwidth of these modes is generally less than $2\kappa$ hence the doublet resonance can be resolved.

The Hamiltonian describing such a WGM doublet resonance is

\begin{multline}
\displaystyle H_{0}=\sum_{k}\omega_{k}\left(a_{k,s}^\dagger a_{k,s}+ a_{k,c}^\dagger a_{k,c}\right) + \\ \sum_k \kappa_{k}\left(a_{k,s}a_{k,c}^\dagger+a_{k,s}^\dagger a_{k,c}\right).
\label{eq:1}
\end{multline}

Here $\omega_k$ is the angular frequency of a WGM with wavenumber $k$, and $a_{k,s}^\dagger$, $a_{k,s}$, $a_{k,c}^\dagger$ and $a_{k,c}$ are the bosonic raising and lowering operators of the {``}$s${''} and {``}$c${''} doublet constituents of this WGM, respectively. The first term in equation (\ref{eq:1}) represents both modes as simple harmonic oscillators (SHOs), while the second term represents the coupling between them, which produces the mode-splitting and doublet appearance. 

A crystal containing dilute concentrations of paramagnetic ion impurities will demonstrate an absorption of energy from these WGMs into the spin angular momentum of the ion{'}s valence electrons if the frequency of the latter transition is tuned (via the Zeeman effect) to be coincident with that of the former. Only WGMs with magnetic field components perpendicular to the applied DC magnetic field will interact in this fashion. This limits the discussion to WGMs that are polarised with a $(H_r,H_\phi,E_z)$ field distribution ({``}WGH{''} modes), since the applied magnetic field in the described case is aligned with the z-axis of the crystal. 

In general, the collective electron spin resonance (ESR) can be considered as an ensemble of independent, non-interacting two level systems (TLSs); and the crystal itself as a paramagnetic material.  In such a case, the ESR, WGM doublet and the interaction between the two can be described by the modified Tavis-Cummings Hamiltonian:

\begin{multline}
\displaystyle H_{TC}=H_{0}+\sum_i\omega_i\sigma_i^+\sigma_i^- + \\\sum_k \sum_i g_{k,s}\left(\sigma_i^-a_{k,s}^\dagger+a_{k,s}\sigma_i^+\right)+\\ \sum_k \sum_i g_{k,c}\left(\sigma_i^-a_{k,c}^\dagger+a_{k,c}\sigma_i^+\right).
\label{eq:2}
\end{multline}

Here, $\omega_i$ is the angular resonant frequency of the $i^{th}$ TLS transition at a particular B-field, and $\sigma_i^+$ and $\sigma_i^-$ are its raising and lowering operators. $g_{k,s}$ ($g_{k,c}$) is the coupling between the spin ensemble and the {``}$s${''} ({``}$c${''}) mode of the WGM with wavenumber $k$.  One of these coupling terms will be set to zero because the ESR can only couple to one resonance of the doublet due to the spin conservation law \cite{gyrotropic}. Note that the choice of which mode will couple ({``}$s${''} or {``}$c${''}) is determined by the sign of the change in spin angular momentum of the TLS transition in question; $\Delta m=\pm 1$.\\

%\noindent If the spin ensemble were in fact capable of forming structured spin waves, the independence of the two-level-systems can no longer be assumed. A structured spin wave, referred to as a {``}magnon{''}, arises from a dipole interaction between neighbouring spins, in which the transition of one spin sets off a {``}chain reaction{''} throughout the ensemble. If the losses of the spin ensemble are small enough such that this {``}chain-reaction{''} of spin flips can travel around the crystal circumference and back to its starting point, a resonant mode can be established; a magnon.
\begin{figure}[t!]
\centering
\includegraphics[width=0.4\textwidth]{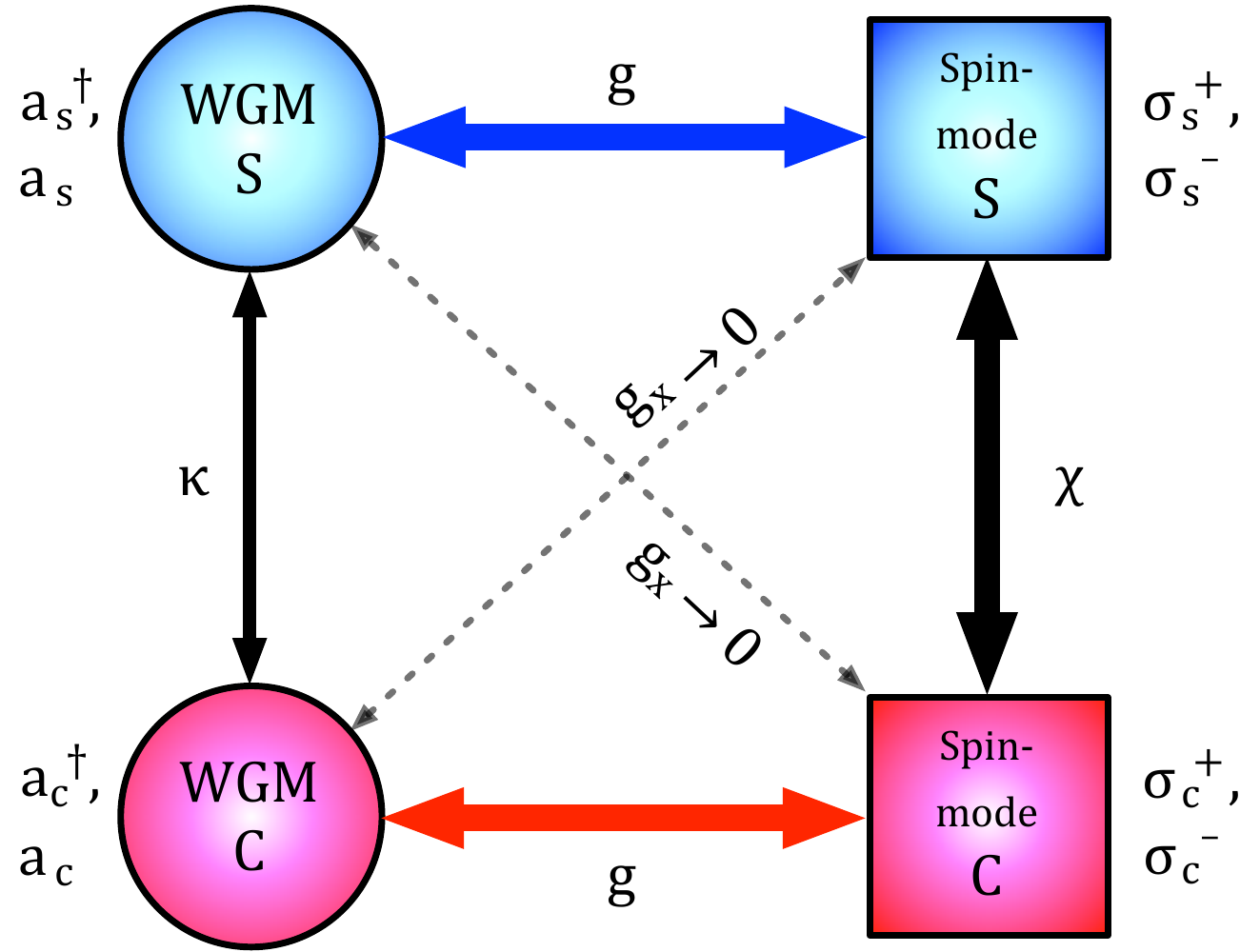}
\caption{(Color online) Four harmonic oscillator model depiction.}
\label{fig:4HO}
\end{figure}

\noindent A spin ensemble can be considered as a classical system of SHOs distributed over a large region of space. For densely packed ensembles interacting with a common cavity mode, these SHOs can be phased relative to each other so that coherent radiation is obtained in a particular direction. This is referred to as superradiance, and occurs when a group of $N$ emitters interact with a common light field in a collective and coherent fashion \cite{dicke}. 

The coherent radiation generated by excited atoms emitting photons, hereon in referred to as the {``}spin-mode{''}, is completely analogous to a photonic WGM, and therefore can exist as a doublet due to backscatterers, in exactly the same way. It will also display the same type of wavenumber orthogonality, doublet orthogonality and a coupling between the two doublet constituents. Spin doublet modes have been previously observed in ferromagnetic YIG samples \cite{YIG}, but never before in a doped sapphire system. In such a scenario, the Hamiltonian describing the interaction between the photonic cavity WGM doublet and the spin-mode doublet would appear as

\begin{multline}
H=H_{0}+\sum_k \omega_k\left(\sigma_{k,s}^+ \sigma_{k,s}^-+\sigma_{k,c}^-\sigma_{k,c}^+ \right)\\+\sum_k g_k\left(\sigma_{k,s}^-a_{k,s}^\dagger +a_{k,s} \sigma_{k,s}^+ +\sigma_{k,c}^- a_{k,c}^\dagger +a_{k,c} \sigma_{k,c}^+ \right)\\+\sum_k \chi_k \left(\sigma_{k,s}^+ \sigma_{k,c}^- + \sigma_{k,c}^+ \sigma_{k,s}^-\right),
\label{eq:3}
\end{multline}

where $\chi$ represents the coupling between the two spin-mode doublet constituents; {``}$s${''} and {``}$c${''}. This Hamiltonian is derived following the treatment of Dicke \cite{dicke} when describing radiation from a gas of large extent. From eq. (\ref{eq:2}); a summation over all modes and TLSs, a transition is made to just the former. The selection rules of such a system \cite{dicke} dictate that only modes with equal wavenumbers, $k$, may interact. In addition to this, the equivalent doublet orthogonality of the spin-modes and WGMs allow for both {``}$s${''}{--}{``}$s${''} and {``}$c${''}{--}{``}$c${''} spin-WGM interactions, but not {``}$s${''}{--}{``}$c${''}. This removes the requirement that one of the spin-mode couplings be set to zero. It is reasonable to assume that the coupling strengths of the two {``}$s${''} modes will be equal to the two {``}$c${''} modes, hence the use of a non-polarisation-specific coupling term, $g_k$. As such, the allowed spin-WGM interactions are described by the third term in eq. (\ref{eq:3}).

The second expression in (\ref{eq:3}) represents the spin-mode doublets as two SHOs, while the final term represents the coupling between them, resulting from imperfections in the crystal; a direct analogue of the last term in eq. (\ref{eq:1}).

Equation (\ref{eq:3}) is represented diagrammatically in FIG. \ref{fig:4HO} for a single value of $k$. It describes a scenario of four SHOs with the allowed linear couplings. This is distinctly different from the case described by eq. (\ref{eq:2}), which would exists as three SHOs \cite{gyrotropic}. \\
% In a circular cavity in which the light wave travels around the boundary, resonant modes can be represented as a linear combination of both clockwise and counterclockwise circularly polarised waves. A WGM doublet is produced by some reflection symmetry breaking, such as back scatterers, lifting the degeneracy of the modes by introducing some coupling between the two counter-propagating waves.  This coupling (and hence the separation of the doublet) is generally on the order of hundreds of kHz, and results in two spatially orthogonal standing waves. Due to WGM{'}s extremely low photon loss ($Q>10^8$), these doublet resonances are clearly resolvable. When the dominant loss mechanism in a cavity is a spin ensemble, it has been shown that one of the waves is a clockwise travelling wave, with right circular polarisation and hence spin angular momentum +1, and the other is the opposite.

%\subsection{Physical Realisation}
\begin{figure}[h!]
\centering
\includegraphics[width=0.4\textwidth]{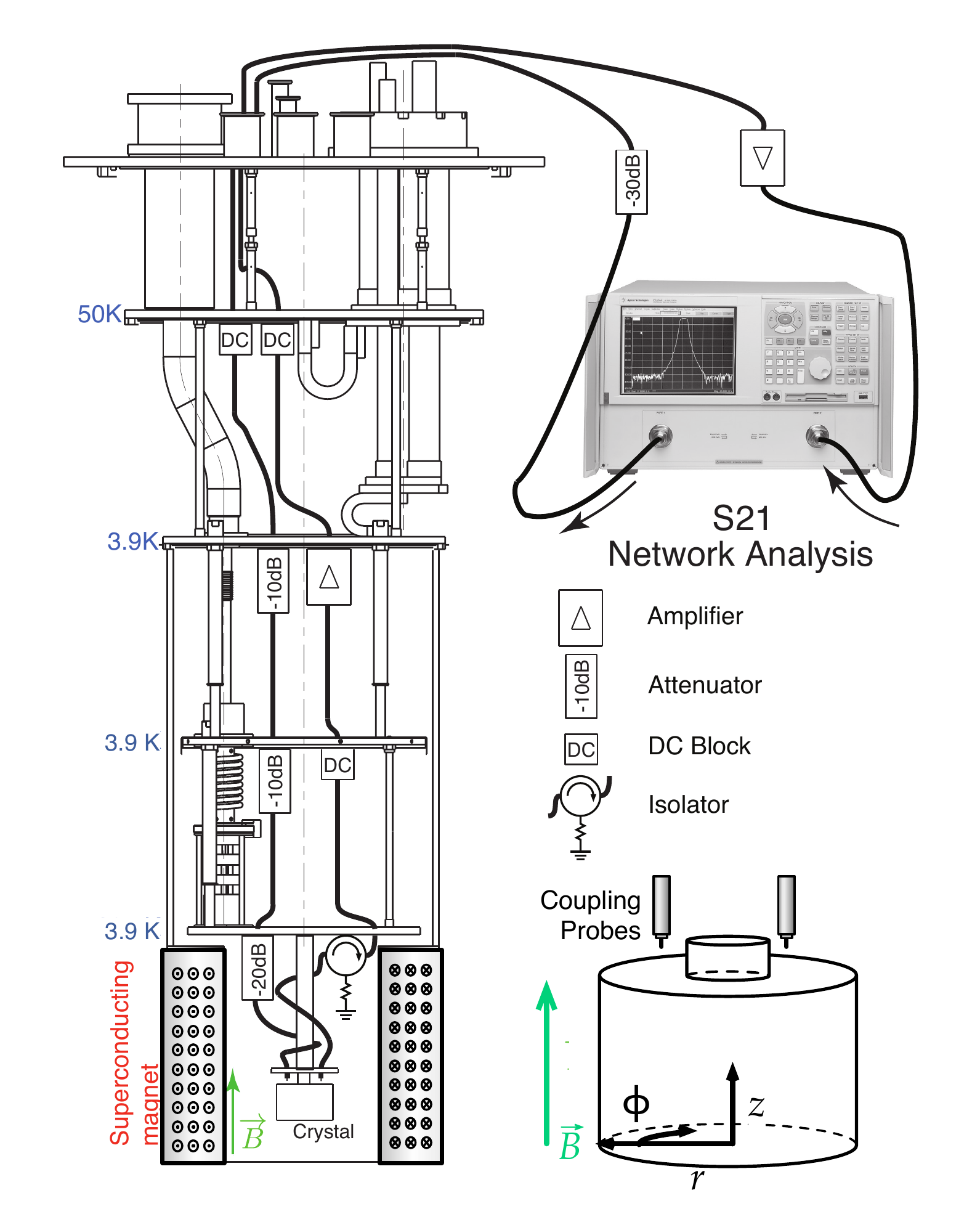}
\caption{(Color online) Experimental setup. The crystal and superconducting magnet are cooled to 4 K in a cryogenic refrigerator, whilst the microwave coupling probes are aligned to excite WGH modes ($E_z$, $H_r$, $H_\phi$ components). Two probes are used to view the relevant modes in transmission (S21) on a Vector Network Analyser (VNA) from which Q factors can be determined as well as the frequency shift caused by coupling to spins.}
\label{fig:exp}
\end{figure}
\noindent The experimental set up is identical to that described by Farr \textit{et al.} \cite{ruby}, however we examine WGMs closer to the zero-field splitting levels of the Cr$^{3+}$ ensemble. The orientation of the crystal, microwave coupling probes and applied DC magnetic field is depicted in FIG. \ref{fig:exp}. Typical ESR parameters for Cr$^{3+}$ ions can be found in \cite{parameters}.  In this paper, we deal with the $\Delta m=\pm1$ transitions; $\left| -3/2\right\rangle \rightarrow \left|-1/2\right\rangle$ and $\left| 3/2\right\rangle \rightarrow \left|1/2\right\rangle$. Both these transitions have a zero-field frequency of 11.447 GHz, and tune in opposite directions as B field is swept ($\Delta m=+1$ increases in frequency with an increase in B field, and vice versa) with $df/dB=\pm g_L \beta$, where $g_L$ is the Land\`{e} g factor and $\beta$ is the Bohr magneton.

\begin{figure}[h!]
\raggedleft
\includegraphics[width=0.4\textwidth]{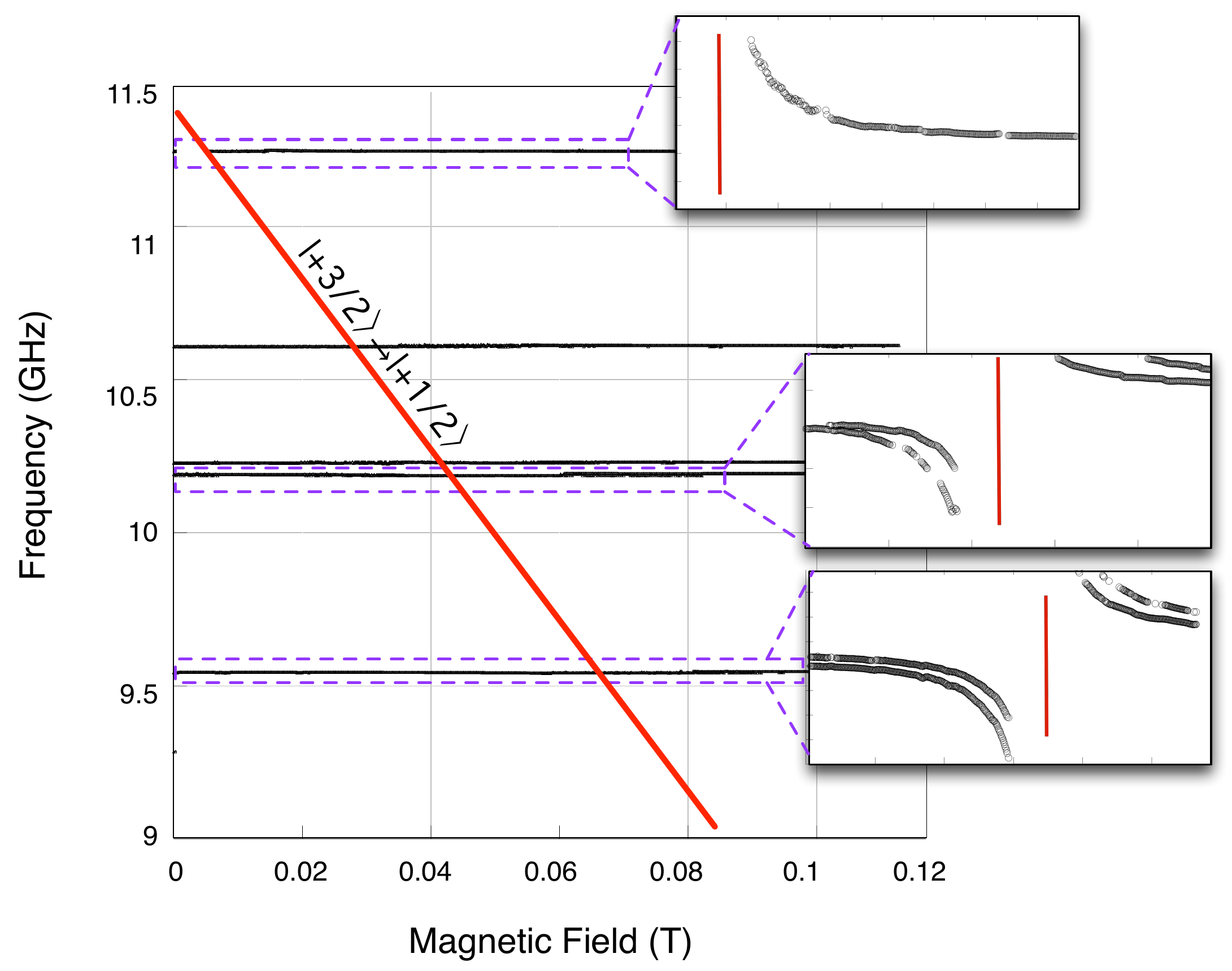}
\caption{(Color online) Spectroscopy results showing the interaction of the generated and pump WGMs with the $\left|+3/2\right\rangle \rightarrow \left|+1/2\right\rangle$ Cr$^{3+}$ electron spin transition as magnetic field is swept. }
\label{fig:transition}
\end{figure}

For example, FIG. \ref{fig:transition} shows the $\Delta m=-1$ transition (in red), as it moves through five distinct WGMs (in black). Each of the data points that make up the black curves represent the position of a resonant peak within a single 8 MHz sweep centred around that particular frequency for that particular magnetic field value. The power incident on the crystal is $P_{inc}=-60$ dBm, which corresponds to a photon occupation number on the order of $10^7$. Far from the intersection of the WGMs and the ESR transition, the black curves in FIG. \ref{fig:transition} represent the frequency location of the WGMs, however when the ESR is tuned such that a particular WGM is within its bandwidth, the black curves represent hybrid spin-WGMs, and an avoided level crossing (ALC) can be observed, as depicted in the inset figures of FIG. \ref{fig:transition}. \\

%The value of this zero-field splitting is $2D=11.447$ GHz $\pm{~}6$ MHz \cite{parameters}. In this paper, we deal with the $\Delta m=\pm1$ transitions; $\left| -3/2\right\rangle \rightarrow \left|-1/2\right\rangle$ and $\left| 3/2\right\rangle \rightarrow \left|1/2\right\rangle$ of the Cr$^{3+}$ ions. A WGM with frequency $2D$ will be coincident with both these transitions at $B=0$. As magnetic field is swept, the transitions are tuned in opposite directions according to the Zeeman effect, according to $f=2D\pm g_L \beta \textbf{B}$, where $g_L$ is the Land\`{e} g-factor and $\beta$ is the Bohr magneton.

%\section{Spin-Photon Interaction}

\begin{figure}[h]
\centering
\includegraphics[width=0.4\textwidth]{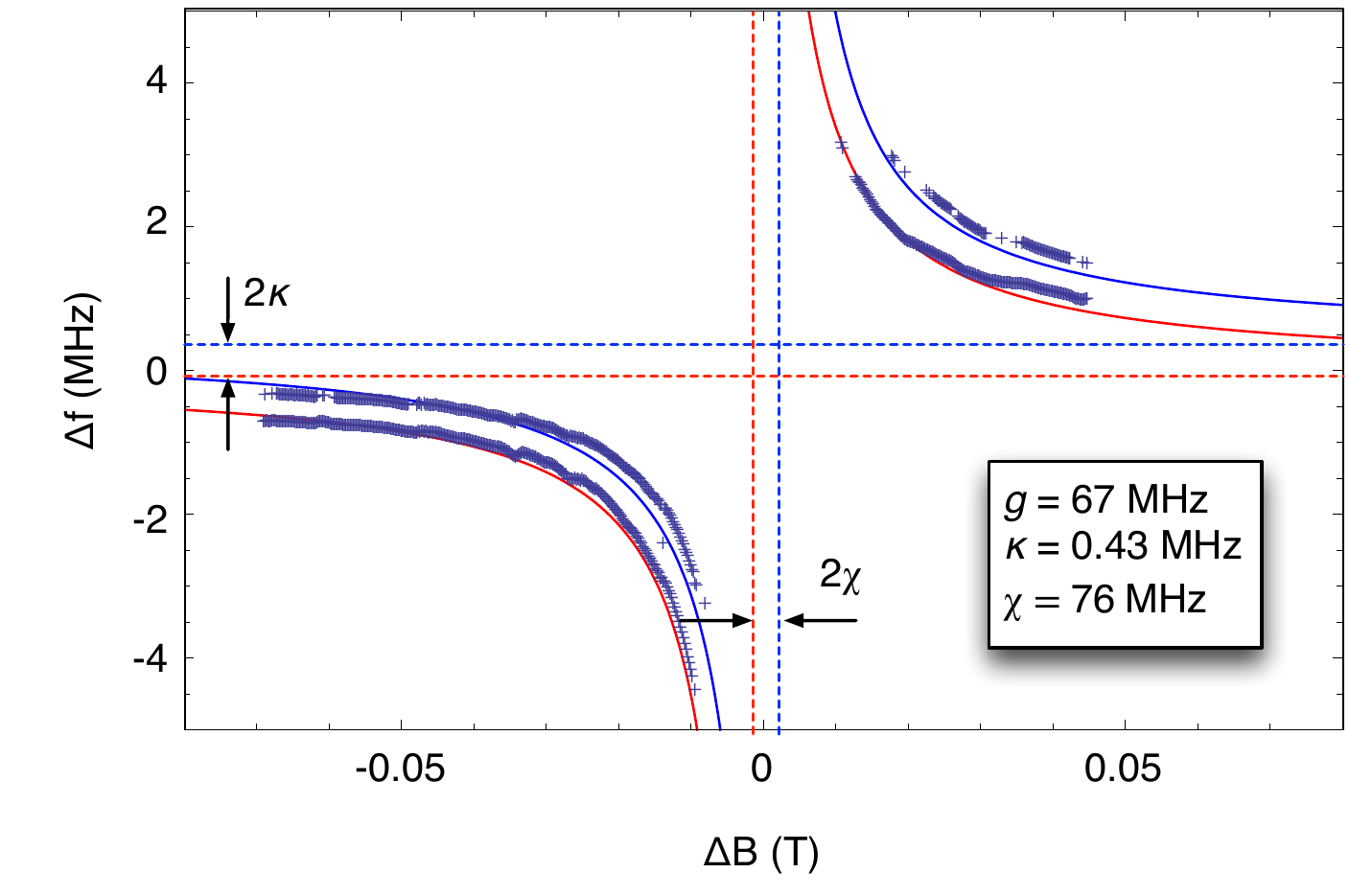}
\caption{(Color online) Interaction of the 9.545 GHz WGM (WGH$_{7,1,1}$) and the Cr$^{3+}$ $\left|+3/2\right\rangle \rightarrow \left|+1/2\right\rangle$ transition at $B=0.68$ T. Asymptotes (dashed) and four SHO model fit (solid).}
\label{fig:ALC}
\end{figure}

%In general, paramagnetic interactions with photonic modes produce an avoided level crossing between the ESR and optical/microwave modes. It has been previously observed in such systems that a gyrotropic response of WGM doublets arises due to the opposite spin angular momentums of the counter-propagating doublet pair. This manifests itself as one interacting and one non-interacting doublet pair, as only photons carrying the correct spin angular momentum can interact. An analytical model for these gyrotropic interactions can produce excellent agreement with observed results. The system is treated as 3 coupled harmonic oscillators (HO), one being the collective spin ensemble and another the non-interacting mode, both of which are coupled to the final HO; the interacting mode. \cite{gyrotropic}
%The modified Tavis-Cummings model (equation (\ref{eq:2})) describes the general case of a WGM doublet and paramagnetic spin ensemble interacting. As has been shown previously, 

\begin{figure}[t!]
\centering
\includegraphics[width=0.4\textwidth]{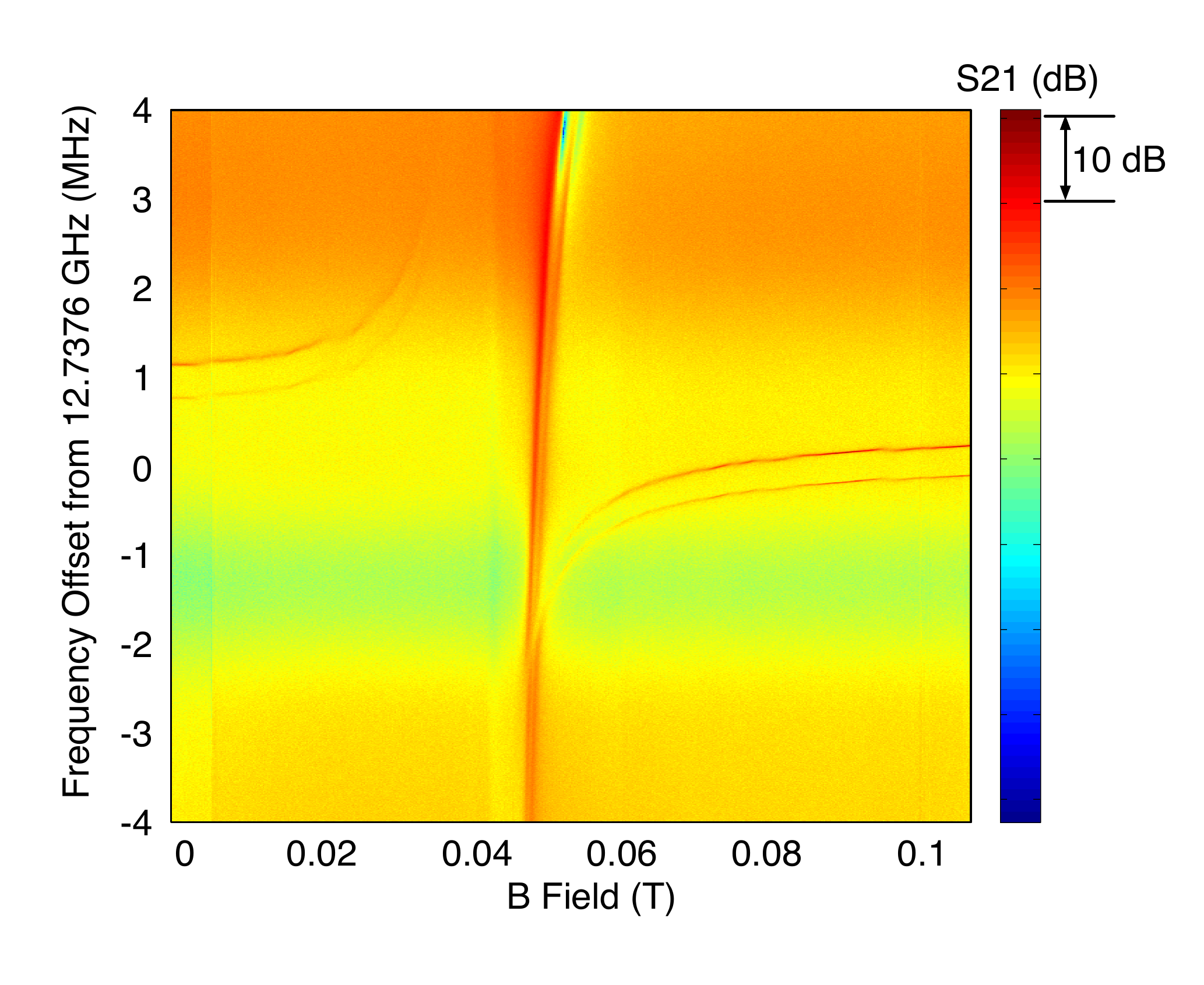}
\caption{(Color online) Interaction of the 12.74 GHz WGM and the Cr$^{3+}$ $\left|-3/2\right\rangle \rightarrow \left|-1/2\right\rangle$ transition.}
\label{fig:cplot}
\end{figure}

\noindent Equation (\ref{eq:2}) predicts a gyrotropic response for the ALC of a WGM doublet and ESR, which may be modelled with great accuracy by three SHOs \cite{gyrotropic}. The ESR spectroscopy results for the ruby crystal in question (FIG. \ref{fig:ALC} and \ref{fig:cplot}) clearly show an absence of this gyrotropic response. We observe that both components of the WGM doublet interact with the spin transition. 

As FIG. \ref{fig:ALC} demonstrates, there are four asymptotes to which the hybrid modes converge. The two horizontal asymptotes of FIG. \ref{fig:ALC} are a standard result of the WGM in question ($f=9.55$ GHz, transverse magnetic mode with 7 azimuthal nodes in $180^\circ$ and 1 radial and axial node {--} WGH$_{7,1,1}$) existing as a doublet. The vertical asymptotes, which in fact depend on B-field just as the red curve in FIG. \ref{fig:transition} (observable if the y-axis scale were broader), confirm the presence of a spin-mode doublet. Their separation is $2\chi$. In the general paramagnetic case of equation (\ref{eq:2}), there would be only one vertical asymptote \cite{gyrotropic}.

The presence of these four asymptotes requires a four SHO model (FIG. \ref{fig:4HO}) to fit the experimental data. Using values of of $g=67$ MHz, $\kappa=0.43$ MHz and $\chi=76$ MHz, a fit is produced which is displayed in FIG. \ref{fig:ALC}. To produce good agreement with the model, it is essential that the cross coupling terms $g_{\times}$ be neglected, or at least be much smaller than the spin-mode couplings $g$ {--} consistent with the allowed terms in the third expression in eq. (\ref{eq:3}). 

%This requirement implies that the two spin-modes interact with the two WGMs independently. This is only possible if the spin ensemble exists as two spatially orthogonal waves, with their spatial distributions matching the WGMs, or at least their spatial distributions must ensure that it is more probable for spin-mode {``}$s${''} to interact with WGM {``}$s${''}, and likewise for {``}$c${''} modes. The spin-mode interaction is therefore actually the interaction between a whispering gallery mode doublet pair with a Òspin wave doubletÓ pair.

The requirement for spatial orthogonality of the spin-modes is again confirmed by FIG. \ref{fig:cplot}. Here, we see the same type of ALC as in FIG. \ref{fig:ALC}. However, we also observe the tail end of another ALC originating at a slightly higher frequency enter the frame. It is the hybrid mode of a higher frequency spin-mode and WGM of a different order. As predicted by the selection rules of such a system, these two doublets simply merge; there is no interaction, due to their different wavenumbers.\\

%Potentially, these four hybrid modes could be coupled as they are formed through an interaction with the same spin ensemble. However, these two doublets simply merge; there is no interaction such as that which would be observed in a {``}superstrong{''} coupling regime. This means that spins involved in the upper ALC do not interact with the spins involved in the lower ALC. They are independent. This cannot be achieved in a paramagnetic (disordered) spin ensemble.  One requires structure to uncouple the two spin-modes, a structure that matches the upper WGM and another that matches the lower WGM. While the spins involved in both interactions may be the same ensemble, the structure of the spin-modes will be different in the same way that two WGMs of different families have different structures.

%\subsection{Spin Losses}

\noindent As WGMs hybridise with a spin-mode (or paramagnetic spin ensembles, for that matter), not only is a frequency shift observable due to the altered magnetic susceptibility of the resonant dielectric, but a change in the now hybrid mode Q factor becomes apparent. This is due to an additional loss mechanism introduced via the coupling to the spin-mode. 

%If the photon-spin coupling were to be in the strong coupling regime, in which a single quantum fully hybridises in equal parts between the two forms, the resultant lifetime of such a particle (its Q-factor) would be equal the average of the two individual, uncoupled particle{'}s lifetimes. In the non-strong coupling regime, such as the one reported here, the Q-factor of the hybrid mode will be equal to the the two individual Q-factors summed in proportion to the level of hybridisation. 

\begin{figure}[t!]
\centering
\includegraphics[width=0.4\textwidth]{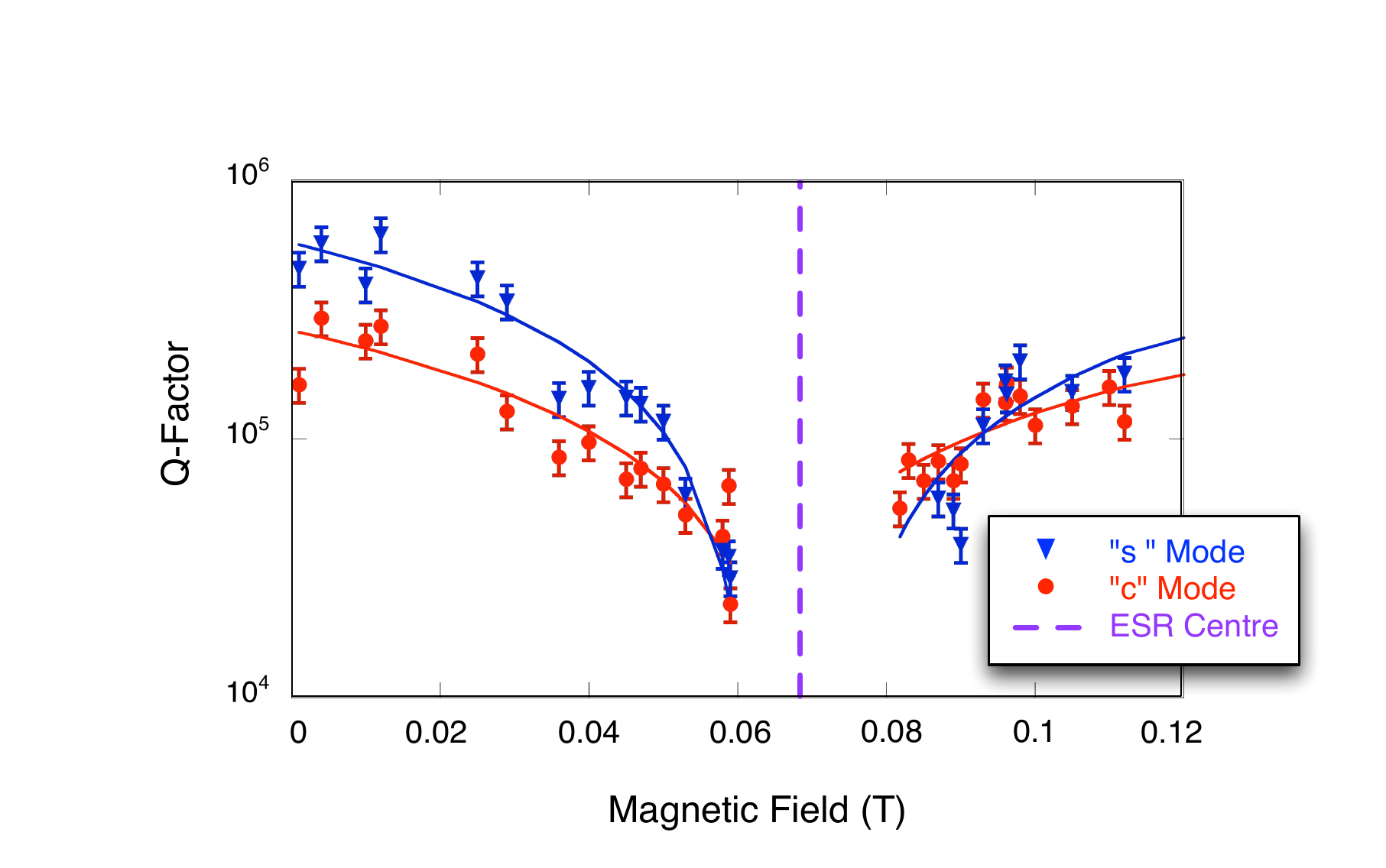}
\caption{(Color online)Q factor of the hybrid {``}$s${''} and {``}$c${''} modes at 9.54 GHz as B-field is swept.}
\label{fig:QvsB}
\end{figure}

The Q factor of the 9.5 GHz mode as B-field is tuned through the ESR centre is plotted in FIG. \ref{fig:QvsB} for both constituents of the mode doublet. As the ESR becomes more closely tuned to the WGM frequency, the mode hybridises to a greater extent and Q factor drops. This is evidence that the losses experienced by spin-modes are greater than those of the photons. Q factors were measured by fitting a Fano resonance line shape \cite{Fano} to the S21 data obtained from the VNA at discrete B-field values.\\

%\subsection{Estimating Impurity Concentration}

\noindent The four HO model fit to the experimental data allows one to make an estimate of the concentration of Cr$^{3+}$  ions in the ruby \cite{ruby}. The concentration of spins participating in the WGM interaction, $n_{part}$, can be calculated using the value of the spin-mode coupling, $g$ and the magnetic filling factor of the WGM in directions perpendicular to the applied DC magnetic field, $\xi$. For the 9.5 GHz WGM, $\xi=0.877$ and was calculated from finite element modelling. $n_{part}$ is calculated to be $8.34\times10^{22}$ ions/m$^3$.

%\begin{equation}
%n_{part}=\frac{4\hbar}{\omega_c \mu_0 \xi}\left(\frac{g}{g_L\beta}\right)^2
%\end{equation}

%where $\mu_0$ is the vacuum permeability and $\xi$ is the magnetic filling factor of the WGM in directions perpendicular to the applied DC magnetic field. 

At finite temperatures, in the continuously driven regime, the number of ions prepared in either the ground ($N_-$) or excited states ($N_+$) of the relevant transition is a function of the thermal distribution of ions and the total number of impurity ions in the crystal, $N_T$. By equating $n_{part} V = N_-$, where $V$ is the total volume of the ruby crystal, one can solve for $N_T$ ($N_T=5.42\times10^{18}$ ions). Given that there are two Al$^{3+}$ ions per unit cell of sapphire that Cr$^{3+}$ can potentially replace, and the volume of the unit cell is that of a trigonal crystal system, the total concentration of ion impurities can be calculated as $N_T$ divided by the total number of potential lattice cites for Cr$^{3+}$ ions to take. A concentration of approximately 40 $ppm$ Cr$^{3+}$ is calculated. This agrees very well with previously measured values for this same crystal, reported as 34 $ppm$ \cite{ruby,parameters}, hence confirming the validity of the four SHO approximation used here, and ergo the conclusions that can be drawn from it.

This concentration is approximately two orders of magnitude larger than the concentration of Fe$^{3+}$ impurities in \cite{gyrotropic} (150 $ppb$), and the value of $g$ is also approximately an order of magnitude larger. In addition, the losses associated with the Cr$^{3+}$ ESR ($\Delta\omega_{spins}/2\pi=9$ MHz) are three times less than those associated with the Fe$^{3+}$ case ($\Delta\omega_{spins}/2\pi=27$ MHz) \cite{gyrotropic,bourhill}. In the present case, $g>\Delta\omega_{spins}$, $\Delta\omega_{WGM}$ ($\Delta\omega_{WGM}/2\pi=Q/f_{res}=1.6$ kHz), satisfying the conditions for strong coupling. It is due to this strong atom-field coupling that superradiance can occur as it is this interaction from which the collective action of the ensemble is derived, and large ion concentrations contribute to this (as $g=g_0\sqrt{N}$). This explains why a superradiant ESR, and hence a four SHO model, is observed in the Cr$^{3+}$ case and not in the previously reported Fe$^{3+}$ case \cite{gyrotropic,bourhill}, which did not satisfy the conditions of strong coupling. 

The tell tale sign of strong coupling (splitting of the resonant cavity mode when the ESR is tuned) is however unobservable due to the loss of coupling between the microwave pump source and WGM when the two are tuned. Because cavity losses are orders of magnitude lower than those associated with the spin ensemble, and the coupling between the transmission line and the WGM, $\beta$, is proportional to $Q$, when the ESR is tuned to the WGM frequency the extra dissipation which is introduced drastically reduces the ability of the transmission line to excite the WGM. Despite the fact that the sum of the spin and cavity mode losses is less than the spin{--}mode coupling, this change in external coupling to the input microwave probe results in the mode vanishing. This is why no resonant mode is observed near the ESR centre as shown in the insets in FIG. \ref{fig:transition}, in FIG. \ref{fig:ALC} about $\Delta B=0$, and hence why no Q factors can be derived in this region in FIG. \ref{fig:QvsB}.\\

\noindent We have described a study regarding the collective interaction of Chromium impurity ions in ruby with crystal photonic WGMs, resulting in the creation of a spin-mode doublet. Theoretical predictions and experimental measurements reveal a set of selection rules that govern the interaction between the spin-modes and WGMs: wavenmuber and azimuthal phase matching. We observe an avoided level crossing between WGM and spin{--}mode doublets in a fashion that can only be described by a four harmonic oscillator model. These results may have important implications for QED experiments dealing with strongly coupled light-matter interactions, as this new model must be used to describe the resulting phenomena. \\

\noindent This work was supported by Australian Research Council grants CE110001013.

%%, and can be approximated by:
%%
%%\begin{equation}
%%\begin{array}{c}
%%\displaystyle N_+=\frac{N_T}{Z}exp\left(-\frac{1}{k_B T}E_{\left| 3/2\right\rangle}\right)\\
%%\\
%%\displaystyle N_-=\frac{N_T}{Z}exp\left(-\frac{1}{k_B T}E_{\left| 1/2\right\rangle}\right),
%%\end{array}
%%\end{equation}
%
%where $Z$ is the corresponding impurity partition function\cite{parameters}, $E$ is the energy of the corresponding state, $N_T$ is the total number of impurity ions, $k_B$ Boltzmann{'}s constant and T the temperature. 

%Given standard values for sapphire of $a=4.75$ \AA{~}and $c=12.982$ \AA, and $g=67$ MHz obtained from fitting the experimental data for the 9.5 GHz mode interaction at B=0.68 T,  


%\bibliography{rubybib}

\begin{thebibliography}{31}%
\makeatletter
\providecommand \@ifxundefined [1]{%
 \@ifx{#1\undefined}
}%
\providecommand \@ifnum [1]{%
 \ifnum #1\expandafter \@firstoftwo
 \else \expandafter \@secondoftwo
 \fi
}%
\providecommand \@ifx [1]{%
 \ifx #1\expandafter \@firstoftwo
 \else \expandafter \@secondoftwo
 \fi
}%
\providecommand \natexlab [1]{#1}%
\providecommand \enquote  [1]{``#1''}%
\providecommand \bibnamefont  [1]{#1}%
\providecommand \bibfnamefont [1]{#1}%
\providecommand \citenamefont [1]{#1}%
\providecommand \href@noop [0]{\@secondoftwo}%
\providecommand \href [0]{\begingroup \@sanitize@url \@href}%
\providecommand \@href[1]{\@@startlink{#1}\@@href}%
\providecommand \@@href[1]{\endgroup#1\@@endlink}%
\providecommand \@sanitize@url [0]{\catcode `\\12\catcode `\$12\catcode
  `\&12\catcode `\#12\catcode `\^12\catcode `\_12\catcode `\%12\relax}%
\providecommand \@@startlink[1]{}%
\providecommand \@@endlink[0]{}%
\providecommand \url  [0]{\begingroup\@sanitize@url \@url }%
\providecommand \@url [1]{\endgroup\@href {#1}{\urlprefix }}%
\providecommand \urlprefix  [0]{URL }%
\providecommand \Eprint [0]{\href }%
\providecommand \doibase [0]{http://dx.doi.org/}%
\providecommand \selectlanguage [0]{\@gobble}%
\providecommand \bibinfo  [0]{\@secondoftwo}%
\providecommand \bibfield  [0]{\@secondoftwo}%
\providecommand \translation [1]{[#1]}%
\providecommand \BibitemOpen [0]{}%
\providecommand \bibitemStop [0]{}%
\providecommand \bibitemNoStop [0]{.\EOS\space}%
\providecommand \EOS [0]{\spacefactor3000\relax}%
\providecommand \BibitemShut  [1]{\csname bibitem#1\endcsname}%
\let\auto@bib@innerbib\@empty
%</preamble>
\bibitem [{\citenamefont {Svidzinsky}\ and\ \citenamefont
  {Chang}(2008)}]{svidzinsky}%
  \BibitemOpen
  \bibfield  {author} {\bibinfo {author} {\bibfnamefont {A.}~\bibnamefont
  {Svidzinsky}}\ and\ \bibinfo {author} {\bibfnamefont {J.-T.}\ \bibnamefont
  {Chang}},\ }\href {\doibase 10.1103/PhysRevA.77.043833} {\bibfield  {journal}
  {\bibinfo  {journal} {Phys. Rev. A}\ }\textbf {\bibinfo {volume} {77}},\
  \bibinfo {pages} {043833} (\bibinfo {year} {2008})}\BibitemShut {NoStop}%
\bibitem [{\citenamefont {Haroche}\ and\ \citenamefont
  {Raimond}(2006)}]{Haroche}%
  \BibitemOpen
  \bibfield  {author} {\bibinfo {author} {\bibfnamefont {S.}~\bibnamefont
  {Haroche}}\ and\ \bibinfo {author} {\bibfnamefont {J.~M.}\ \bibnamefont
  {Raimond}},\ }\href {http://cds.cern.ch/record/993568} {\emph {\bibinfo
  {title} {{Exploring the Quantum: Atoms, Cavities, and Photons}}}}\ (\bibinfo
  {publisher} {Oxford Univ. Press},\ \bibinfo {address} {Oxford},\ \bibinfo
  {year} {2006})\BibitemShut {NoStop}%
\bibitem [{\citenamefont {Tian}\ \emph {et~al.}(2004)\citenamefont {Tian},
  \citenamefont {Rabl}, \citenamefont {Blatt},\ and\ \citenamefont
  {Zoller}}]{optmic5}%
  \BibitemOpen
  \bibfield  {author} {\bibinfo {author} {\bibfnamefont {L.}~\bibnamefont
  {Tian}}, \bibinfo {author} {\bibfnamefont {P.}~\bibnamefont {Rabl}}, \bibinfo
  {author} {\bibfnamefont {R.}~\bibnamefont {Blatt}}, \ and\ \bibinfo {author}
  {\bibfnamefont {P.}~\bibnamefont {Zoller}},\ }\href {\doibase
  10.1103/PhysRevLett.92.247902} {\bibfield  {journal} {\bibinfo  {journal}
  {Phys. Rev. Lett.}\ }\textbf {\bibinfo {volume} {92}},\ \bibinfo {pages}
  {247902} (\bibinfo {year} {2004})}\BibitemShut {NoStop}%
\bibitem [{\citenamefont {Probst}\ \emph {et~al.}(2014)\citenamefont {Probst},
  \citenamefont {Tkal\ifmmode~\check{c}\else \v{c}\fi{}ec}, \citenamefont
  {Rotzinger}, \citenamefont {Rieger}, \citenamefont {Le~Floch}, \citenamefont
  {Goryachev}, \citenamefont {Tobar}, \citenamefont {Ustinov},\ and\
  \citenamefont {Bushev}}]{optmic1}%
  \BibitemOpen
  \bibfield  {author} {\bibinfo {author} {\bibfnamefont {S.}~\bibnamefont
  {Probst}}, \bibinfo {author} {\bibfnamefont {A.}~\bibnamefont
  {Tkal\ifmmode~\check{c}\else \v{c}\fi{}ec}}, \bibinfo {author} {\bibfnamefont
  {H.}~\bibnamefont {Rotzinger}}, \bibinfo {author} {\bibfnamefont
  {D.}~\bibnamefont {Rieger}}, \bibinfo {author} {\bibfnamefont {J.-M.}\
  \bibnamefont {Le~Floch}}, \bibinfo {author} {\bibfnamefont {M.}~\bibnamefont
  {Goryachev}}, \bibinfo {author} {\bibfnamefont {M.~E.}\ \bibnamefont
  {Tobar}}, \bibinfo {author} {\bibfnamefont {A.~V.}\ \bibnamefont {Ustinov}},
  \ and\ \bibinfo {author} {\bibfnamefont {P.~A.}\ \bibnamefont {Bushev}},\
  }\href {\doibase 10.1103/PhysRevB.90.100404} {\bibfield  {journal} {\bibinfo
  {journal} {Phys. Rev. B}\ }\textbf {\bibinfo {volume} {90}},\ \bibinfo
  {pages} {100404} (\bibinfo {year} {2014})}\BibitemShut {NoStop}%
\bibitem [{\citenamefont {Probst}\ \emph {et~al.}(2013)\citenamefont {Probst},
  \citenamefont {Rotzinger}, \citenamefont {W\"unsch}, \citenamefont {Jung},
  \citenamefont {Jerger}, \citenamefont {Siegel}, \citenamefont {Ustinov},\
  and\ \citenamefont {Bushev}}]{optmic4}%
  \BibitemOpen
  \bibfield  {author} {\bibinfo {author} {\bibfnamefont {S.}~\bibnamefont
  {Probst}}, \bibinfo {author} {\bibfnamefont {H.}~\bibnamefont {Rotzinger}},
  \bibinfo {author} {\bibfnamefont {S.}~\bibnamefont {W\"unsch}}, \bibinfo
  {author} {\bibfnamefont {P.}~\bibnamefont {Jung}}, \bibinfo {author}
  {\bibfnamefont {M.}~\bibnamefont {Jerger}}, \bibinfo {author} {\bibfnamefont
  {M.}~\bibnamefont {Siegel}}, \bibinfo {author} {\bibfnamefont {A.~V.}\
  \bibnamefont {Ustinov}}, \ and\ \bibinfo {author} {\bibfnamefont {P.~A.}\
  \bibnamefont {Bushev}},\ }\href {\doibase 10.1103/PhysRevLett.110.157001}
  {\bibfield  {journal} {\bibinfo  {journal} {Phys. Rev. Lett.}\ }\textbf
  {\bibinfo {volume} {110}},\ \bibinfo {pages} {157001} (\bibinfo {year}
  {2013})}\BibitemShut {NoStop}%
\bibitem [{\citenamefont {Imamo\ifmmode~\breve{g}\else
  \u{g}\fi{}lu}(2009)}]{optmic2}%
  \BibitemOpen
  \bibfield  {author} {\bibinfo {author} {\bibfnamefont {A.}~\bibnamefont
  {Imamo\ifmmode~\breve{g}\else \u{g}\fi{}lu}},\ }\href {\doibase
  10.1103/PhysRevLett.102.083602} {\bibfield  {journal} {\bibinfo  {journal}
  {Phys. Rev. Lett.}\ }\textbf {\bibinfo {volume} {102}},\ \bibinfo {pages}
  {083602} (\bibinfo {year} {2009})}\BibitemShut {NoStop}%
\bibitem [{\citenamefont {Schuster}\ \emph {et~al.}(2010)\citenamefont
  {Schuster}, \citenamefont {Sears}, \citenamefont {Ginossar}, \citenamefont
  {DiCarlo}, \citenamefont {Frunzio}, \citenamefont {Morton}, \citenamefont
  {Wu}, \citenamefont {Briggs}, \citenamefont {Buckley}, \citenamefont
  {Awschalom},\ and\ \citenamefont {Schoelkopf}}]{optmic3}%
  \BibitemOpen
  \bibfield  {author} {\bibinfo {author} {\bibfnamefont {D.~I.}\ \bibnamefont
  {Schuster}}, \bibinfo {author} {\bibfnamefont {A.~P.}\ \bibnamefont {Sears}},
  \bibinfo {author} {\bibfnamefont {E.}~\bibnamefont {Ginossar}}, \bibinfo
  {author} {\bibfnamefont {L.}~\bibnamefont {DiCarlo}}, \bibinfo {author}
  {\bibfnamefont {L.}~\bibnamefont {Frunzio}}, \bibinfo {author} {\bibfnamefont
  {J.~J.~L.}\ \bibnamefont {Morton}}, \bibinfo {author} {\bibfnamefont
  {H.}~\bibnamefont {Wu}}, \bibinfo {author} {\bibfnamefont {G.~A.~D.}\
  \bibnamefont {Briggs}}, \bibinfo {author} {\bibfnamefont {B.~B.}\
  \bibnamefont {Buckley}}, \bibinfo {author} {\bibfnamefont {D.~D.}\
  \bibnamefont {Awschalom}}, \ and\ \bibinfo {author} {\bibfnamefont {R.~J.}\
  \bibnamefont {Schoelkopf}},\ }\href {\doibase 10.1103/PhysRevLett.105.140501}
  {\bibfield  {journal} {\bibinfo  {journal} {Phys. Rev. Lett.}\ }\textbf
  {\bibinfo {volume} {105}},\ \bibinfo {pages} {140501} (\bibinfo {year}
  {2010})}\BibitemShut {NoStop}%
\bibitem [{\citenamefont {Dicke}(1954)}]{dicke}%
  \BibitemOpen
  \bibfield  {author} {\bibinfo {author} {\bibfnamefont {R.~H.}\ \bibnamefont
  {Dicke}},\ }\href@noop {} {\bibfield  {journal} {\bibinfo  {journal}
  {Physical Review}\ }\textbf {\bibinfo {volume} {93}},\ \bibinfo {pages} {99}
  (\bibinfo {year} {1954})}\BibitemShut {NoStop}%
\bibitem [{\citenamefont {Scully}\ and\ \citenamefont
  {Svidzinsky}(2009)}]{Scully}%
  \BibitemOpen
  \bibfield  {author} {\bibinfo {author} {\bibfnamefont {M.~O.}\ \bibnamefont
  {Scully}}\ and\ \bibinfo {author} {\bibfnamefont {A.~A.}\ \bibnamefont
  {Svidzinsky}},\ }\href {\doibase 10.1126/science.1176695} {\bibfield
  {journal} {\bibinfo  {journal} {Science}\ }\textbf {\bibinfo {volume}
  {325}},\ \bibinfo {pages} {1510} (\bibinfo {year} {2009})}\BibitemShut
  {NoStop}%
\bibitem [{\citenamefont {Skribanowitz}\ \emph {et~al.}(1973)\citenamefont
  {Skribanowitz}, \citenamefont {Herman}, \citenamefont {MacGillivray},\ and\
  \citenamefont {Feld}}]{HF}%
  \BibitemOpen
  \bibfield  {author} {\bibinfo {author} {\bibfnamefont {N.}~\bibnamefont
  {Skribanowitz}}, \bibinfo {author} {\bibfnamefont {I.~P.}\ \bibnamefont
  {Herman}}, \bibinfo {author} {\bibfnamefont {J.~C.}\ \bibnamefont
  {MacGillivray}}, \ and\ \bibinfo {author} {\bibfnamefont {M.~S.}\
  \bibnamefont {Feld}},\ }\href {\doibase 10.1103/PhysRevLett.30.309}
  {\bibfield  {journal} {\bibinfo  {journal} {Phys. Rev. Lett.}\ }\textbf
  {\bibinfo {volume} {30}},\ \bibinfo {pages} {309} (\bibinfo {year}
  {1973})}\BibitemShut {NoStop}%
\bibitem [{\citenamefont {Xia}\ \emph {et~al.}(2012)\citenamefont {Xia},
  \citenamefont {Svidzinsky}, \citenamefont {Yuan}, \citenamefont {Lu},
  \citenamefont {Suckewer},\ and\ \citenamefont {Scully}}]{He}%
  \BibitemOpen
  \bibfield  {author} {\bibinfo {author} {\bibfnamefont {H.}~\bibnamefont
  {Xia}}, \bibinfo {author} {\bibfnamefont {A.~A.}\ \bibnamefont {Svidzinsky}},
  \bibinfo {author} {\bibfnamefont {L.}~\bibnamefont {Yuan}}, \bibinfo {author}
  {\bibfnamefont {C.}~\bibnamefont {Lu}}, \bibinfo {author} {\bibfnamefont
  {S.}~\bibnamefont {Suckewer}}, \ and\ \bibinfo {author} {\bibfnamefont
  {M.~O.}\ \bibnamefont {Scully}},\ }\href {\doibase
  10.1103/PhysRevLett.109.093604} {\bibfield  {journal} {\bibinfo  {journal}
  {Phys. Rev. Lett.}\ }\textbf {\bibinfo {volume} {109}},\ \bibinfo {pages}
  {093604} (\bibinfo {year} {2012})}\BibitemShut {NoStop}%
\bibitem [{\citenamefont {Sadler}\ \emph {et~al.}(2007)\citenamefont {Sadler},
  \citenamefont {Higbie}, \citenamefont {Leslie}, \citenamefont
  {Vengalattore},\ and\ \citenamefont {Stamper-Kurn}}]{Rb}%
  \BibitemOpen
  \bibfield  {author} {\bibinfo {author} {\bibfnamefont {L.~E.}\ \bibnamefont
  {Sadler}}, \bibinfo {author} {\bibfnamefont {J.~M.}\ \bibnamefont {Higbie}},
  \bibinfo {author} {\bibfnamefont {S.~R.}\ \bibnamefont {Leslie}}, \bibinfo
  {author} {\bibfnamefont {M.}~\bibnamefont {Vengalattore}}, \ and\ \bibinfo
  {author} {\bibfnamefont {D.~M.}\ \bibnamefont {Stamper-Kurn}},\ }\href
  {\doibase 10.1103/PhysRevLett.98.110401} {\bibfield  {journal} {\bibinfo
  {journal} {Phys. Rev. Lett.}\ }\textbf {\bibinfo {volume} {98}},\ \bibinfo
  {pages} {110401} (\bibinfo {year} {2007})}\BibitemShut {NoStop}%
\bibitem [{\citenamefont {Crubellier}\ \emph {et~al.}(1978)\citenamefont
  {Crubellier}, \citenamefont {Liberman},\ and\ \citenamefont {Pillet}}]{Rb2}%
  \BibitemOpen
  \bibfield  {author} {\bibinfo {author} {\bibfnamefont {A.}~\bibnamefont
  {Crubellier}}, \bibinfo {author} {\bibfnamefont {S.}~\bibnamefont
  {Liberman}}, \ and\ \bibinfo {author} {\bibfnamefont {P.}~\bibnamefont
  {Pillet}},\ }\href {\doibase 10.1103/PhysRevLett.41.1237} {\bibfield
  {journal} {\bibinfo  {journal} {Phys. Rev. Lett.}\ }\textbf {\bibinfo
  {volume} {41}},\ \bibinfo {pages} {1237} (\bibinfo {year}
  {1978})}\BibitemShut {NoStop}%
\bibitem [{\citenamefont {Gross}\ \emph {et~al.}(1976)\citenamefont {Gross},
  \citenamefont {Fabre}, \citenamefont {Pillet},\ and\ \citenamefont
  {Haroche}}]{Na}%
  \BibitemOpen
  \bibfield  {author} {\bibinfo {author} {\bibfnamefont {M.}~\bibnamefont
  {Gross}}, \bibinfo {author} {\bibfnamefont {C.}~\bibnamefont {Fabre}},
  \bibinfo {author} {\bibfnamefont {P.}~\bibnamefont {Pillet}}, \ and\ \bibinfo
  {author} {\bibfnamefont {S.}~\bibnamefont {Haroche}},\ }\href {\doibase
  10.1103/PhysRevLett.36.1035} {\bibfield  {journal} {\bibinfo  {journal}
  {Phys. Rev. Lett.}\ }\textbf {\bibinfo {volume} {36}},\ \bibinfo {pages}
  {1035} (\bibinfo {year} {1976})}\BibitemShut {NoStop}%
\bibitem [{\citenamefont {Meinardi}\ \emph {et~al.}(2003)\citenamefont
  {Meinardi}, \citenamefont {Cerminara}, \citenamefont {Sassella},
  \citenamefont {Bonifacio},\ and\ \citenamefont {Tubino}}]{Hagg}%
  \BibitemOpen
  \bibfield  {author} {\bibinfo {author} {\bibfnamefont {F.}~\bibnamefont
  {Meinardi}}, \bibinfo {author} {\bibfnamefont {M.}~\bibnamefont {Cerminara}},
  \bibinfo {author} {\bibfnamefont {A.}~\bibnamefont {Sassella}}, \bibinfo
  {author} {\bibfnamefont {R.}~\bibnamefont {Bonifacio}}, \ and\ \bibinfo
  {author} {\bibfnamefont {R.}~\bibnamefont {Tubino}},\ }\href {\doibase
  10.1103/PhysRevLett.91.247401} {\bibfield  {journal} {\bibinfo  {journal}
  {Phys. Rev. Lett.}\ }\textbf {\bibinfo {volume} {91}},\ \bibinfo {pages}
  {247401} (\bibinfo {year} {2003})}\BibitemShut {NoStop}%
\bibitem [{\citenamefont {Lim}\ \emph {et~al.}(2004)\citenamefont {Lim},
  \citenamefont {Bjorklund}, \citenamefont {Spano},\ and\ \citenamefont
  {Bardeen}}]{tetracene}%
  \BibitemOpen
  \bibfield  {author} {\bibinfo {author} {\bibfnamefont {S.-H.}\ \bibnamefont
  {Lim}}, \bibinfo {author} {\bibfnamefont {T.~G.}\ \bibnamefont {Bjorklund}},
  \bibinfo {author} {\bibfnamefont {F.~C.}\ \bibnamefont {Spano}}, \ and\
  \bibinfo {author} {\bibfnamefont {C.~J.}\ \bibnamefont {Bardeen}},\ }\href
  {\doibase 10.1103/PhysRevLett.92.107402} {\bibfield  {journal} {\bibinfo
  {journal} {Phys. Rev. Lett.}\ }\textbf {\bibinfo {volume} {92}},\ \bibinfo
  {pages} {107402} (\bibinfo {year} {2004})}\BibitemShut {NoStop}%
\bibitem [{\citenamefont {Frolov}\ \emph {et~al.}(1997)\citenamefont {Frolov},
  \citenamefont {Gellermann}, \citenamefont {Ozaki}, \citenamefont {Yoshino},\
  and\ \citenamefont {Vardeny}}]{poly}%
  \BibitemOpen
  \bibfield  {author} {\bibinfo {author} {\bibfnamefont {S.~V.}\ \bibnamefont
  {Frolov}}, \bibinfo {author} {\bibfnamefont {W.}~\bibnamefont {Gellermann}},
  \bibinfo {author} {\bibfnamefont {M.}~\bibnamefont {Ozaki}}, \bibinfo
  {author} {\bibfnamefont {K.}~\bibnamefont {Yoshino}}, \ and\ \bibinfo
  {author} {\bibfnamefont {Z.~V.}\ \bibnamefont {Vardeny}},\ }\href {\doibase
  10.1103/PhysRevLett.78.729} {\bibfield  {journal} {\bibinfo  {journal} {Phys.
  Rev. Lett.}\ }\textbf {\bibinfo {volume} {78}},\ \bibinfo {pages} {729}
  (\bibinfo {year} {1997})}\BibitemShut {NoStop}%
\bibitem [{\citenamefont {Greiner}\ \emph {et~al.}(2000)\citenamefont
  {Greiner}, \citenamefont {Boggs},\ and\ \citenamefont {Mossberg}}]{YAG}%
  \BibitemOpen
  \bibfield  {author} {\bibinfo {author} {\bibfnamefont {C.}~\bibnamefont
  {Greiner}}, \bibinfo {author} {\bibfnamefont {B.}~\bibnamefont {Boggs}}, \
  and\ \bibinfo {author} {\bibfnamefont {T.~W.}\ \bibnamefont {Mossberg}},\
  }\href {\doibase 10.1103/PhysRevLett.85.3793} {\bibfield  {journal} {\bibinfo
   {journal} {Phys. Rev. Lett.}\ }\textbf {\bibinfo {volume} {85}},\ \bibinfo
  {pages} {3793} (\bibinfo {year} {2000})}\BibitemShut {NoStop}%
\bibitem [{\citenamefont {Kondo}\ \emph
  {et~al.}(2004{\natexlab{a}})\citenamefont {Kondo}, \citenamefont {Nakagawa},
  \citenamefont {Saito},\ and\ \citenamefont {Asada}}]{crystal}%
  \BibitemOpen
  \bibfield  {author} {\bibinfo {author} {\bibfnamefont {S.}~\bibnamefont
  {Kondo}}, \bibinfo {author} {\bibfnamefont {H.}~\bibnamefont {Nakagawa}},
  \bibinfo {author} {\bibfnamefont {T.}~\bibnamefont {Saito}}, \ and\ \bibinfo
  {author} {\bibfnamefont {H.}~\bibnamefont {Asada}},\ }\href {\doibase
  http://dx.doi.org/10.1016/j.cap.2003.11.063} {\bibfield  {journal} {\bibinfo
  {journal} {Current Applied Physics}\ }\textbf {\bibinfo {volume} {4}},\
  \bibinfo {pages} {439 } (\bibinfo {year} {2004}{\natexlab{a}})}\BibitemShut
  {NoStop}%
\bibitem [{\citenamefont {Kondo}\ \emph
  {et~al.}(2004{\natexlab{b}})\citenamefont {Kondo}, \citenamefont {Suzuki},
  \citenamefont {Saito}, \citenamefont {Asada},\ and\ \citenamefont
  {Nakagawa}}]{crystal2}%
  \BibitemOpen
  \bibfield  {author} {\bibinfo {author} {\bibfnamefont {S.}~\bibnamefont
  {Kondo}}, \bibinfo {author} {\bibfnamefont {K.}~\bibnamefont {Suzuki}},
  \bibinfo {author} {\bibfnamefont {T.}~\bibnamefont {Saito}}, \bibinfo
  {author} {\bibfnamefont {H.}~\bibnamefont {Asada}}, \ and\ \bibinfo {author}
  {\bibfnamefont {H.}~\bibnamefont {Nakagawa}},\ }\href {\doibase
  10.1103/PhysRevB.70.205322} {\bibfield  {journal} {\bibinfo  {journal} {Phys.
  Rev. B}\ }\textbf {\bibinfo {volume} {70}},\ \bibinfo {pages} {205322}
  (\bibinfo {year} {2004}{\natexlab{b}})}\BibitemShut {NoStop}%
\bibitem [{\citenamefont {Barenco}\ \emph {et~al.}(1995)\citenamefont
  {Barenco}, \citenamefont {Deutsch}, \citenamefont {Ekert},\ and\
  \citenamefont {Jozsa}}]{dot1}%
  \BibitemOpen
  \bibfield  {author} {\bibinfo {author} {\bibfnamefont {A.}~\bibnamefont
  {Barenco}}, \bibinfo {author} {\bibfnamefont {D.}~\bibnamefont {Deutsch}},
  \bibinfo {author} {\bibfnamefont {A.}~\bibnamefont {Ekert}}, \ and\ \bibinfo
  {author} {\bibfnamefont {R.}~\bibnamefont {Jozsa}},\ }\href {\doibase
  10.1103/PhysRevLett.74.4083} {\bibfield  {journal} {\bibinfo  {journal}
  {Phys. Rev. Lett.}\ }\textbf {\bibinfo {volume} {74}},\ \bibinfo {pages}
  {4083} (\bibinfo {year} {1995})}\BibitemShut {NoStop}%
\bibitem [{\citenamefont {Bayer}\ \emph {et~al.}(2001)\citenamefont {Bayer},
  \citenamefont {Hawrylak}, \citenamefont {Hinzer}, \citenamefont {Fafard},
  \citenamefont {Korkusinski}, \citenamefont {Wasilewski}, \citenamefont
  {Stern},\ and\ \citenamefont {Forchel}}]{dot2}%
  \BibitemOpen
  \bibfield  {author} {\bibinfo {author} {\bibfnamefont {M.}~\bibnamefont
  {Bayer}}, \bibinfo {author} {\bibfnamefont {P.}~\bibnamefont {Hawrylak}},
  \bibinfo {author} {\bibfnamefont {K.}~\bibnamefont {Hinzer}}, \bibinfo
  {author} {\bibfnamefont {S.}~\bibnamefont {Fafard}}, \bibinfo {author}
  {\bibfnamefont {M.}~\bibnamefont {Korkusinski}}, \bibinfo {author}
  {\bibfnamefont {Z.~R.}\ \bibnamefont {Wasilewski}}, \bibinfo {author}
  {\bibfnamefont {O.}~\bibnamefont {Stern}}, \ and\ \bibinfo {author}
  {\bibfnamefont {A.}~\bibnamefont {Forchel}},\ }\href {\doibase
  10.1126/science.291.5503.451} {\bibfield  {journal} {\bibinfo  {journal}
  {Science}\ }\textbf {\bibinfo {volume} {291}},\ \bibinfo {pages} {451}
  (\bibinfo {year} {2001})}\BibitemShut {NoStop}%
\bibitem [{\citenamefont {Mlynek}\ \emph {et~al.}(2014)\citenamefont {Mlynek},
  \citenamefont {Abdumalikov}, \citenamefont {Eichler},\ and\ \citenamefont
  {Wallraff}}]{dot3}%
  \BibitemOpen
  \bibfield  {author} {\bibinfo {author} {\bibfnamefont {J.~A.}\ \bibnamefont
  {Mlynek}}, \bibinfo {author} {\bibfnamefont {A.~A.}\ \bibnamefont
  {Abdumalikov}}, \bibinfo {author} {\bibfnamefont {C.}~\bibnamefont
  {Eichler}}, \ and\ \bibinfo {author} {\bibfnamefont {A.}~\bibnamefont
  {Wallraff}},\ }\href {http://dx.doi.org/10.1038/ncomms6186} {\bibfield
  {journal} {\bibinfo  {journal} {Nat Commun}\ }\textbf {\bibinfo {volume} {5}}
  (\bibinfo {year} {2014})}\BibitemShut {NoStop}%
\bibitem [{\citenamefont {Tobar}(2005)}]{tobar1}%
  \BibitemOpen
  \bibfield  {author} {\bibinfo {author} {\bibfnamefont {M.~E.}\ \bibnamefont
  {Tobar}},\ }\href {http://stacks.iop.org/0026-1394/42/i=2/a=007} {\bibfield
  {journal} {\bibinfo  {journal} {Metrologia}\ }\textbf {\bibinfo {volume}
  {42}},\ \bibinfo {pages} {129} (\bibinfo {year} {2005})}\BibitemShut
  {NoStop}%
\bibitem [{\citenamefont {Farr}\ \emph {et~al.}(2014)\citenamefont {Farr},
  \citenamefont {Goryachev}, \citenamefont {Creedon},\ and\ \citenamefont
  {Tobar}}]{ruby}%
  \BibitemOpen
  \bibfield  {author} {\bibinfo {author} {\bibfnamefont {W.~G.}\ \bibnamefont
  {Farr}}, \bibinfo {author} {\bibfnamefont {M.}~\bibnamefont {Goryachev}},
  \bibinfo {author} {\bibfnamefont {D.~L.}\ \bibnamefont {Creedon}}, \ and\
  \bibinfo {author} {\bibfnamefont {M.~E.}\ \bibnamefont {Tobar}},\ }\href
  {\doibase 10.1103/PhysRevB.90.054409} {\bibfield  {journal} {\bibinfo
  {journal} {Phys. Rev. B}\ }\textbf {\bibinfo {volume} {90}},\ \bibinfo
  {pages} {054409} (\bibinfo {year} {2014})}\BibitemShut {NoStop}%
\bibitem [{\citenamefont {Bourgeois}\ and\ \citenamefont
  {Giordano}(2005)}]{backscatter}%
  \BibitemOpen
  \bibfield  {author} {\bibinfo {author} {\bibfnamefont {P.}~\bibnamefont
  {Bourgeois}}\ and\ \bibinfo {author} {\bibfnamefont {V.}~\bibnamefont
  {Giordano}},\ }\href {\doibase 10.1109/TMTT.2005.855145} {\bibfield
  {journal} {\bibinfo  {journal} {Microwave Theory and Techniques, IEEE
  Transactions on}\ }\textbf {\bibinfo {volume} {53}},\ \bibinfo {pages} {3185}
  (\bibinfo {year} {2005})}\BibitemShut {NoStop}%
\bibitem [{\citenamefont {Goryachev}\ \emph
  {et~al.}(2014{\natexlab{a}})\citenamefont {Goryachev}, \citenamefont {Farr},
  \citenamefont {Creedon},\ and\ \citenamefont {Tobar}}]{gyrotropic}%
  \BibitemOpen
  \bibfield  {author} {\bibinfo {author} {\bibfnamefont {M.}~\bibnamefont
  {Goryachev}}, \bibinfo {author} {\bibfnamefont {W.~G.}\ \bibnamefont {Farr}},
  \bibinfo {author} {\bibfnamefont {D.~L.}\ \bibnamefont {Creedon}}, \ and\
  \bibinfo {author} {\bibfnamefont {M.~E.}\ \bibnamefont {Tobar}},\ }\href
  {\doibase 10.1103/PhysRevB.89.224407} {\bibfield  {journal} {\bibinfo
  {journal} {Phys. Rev. B}\ }\textbf {\bibinfo {volume} {89}},\ \bibinfo
  {pages} {224407} (\bibinfo {year} {2014}{\natexlab{a}})}\BibitemShut
  {NoStop}%
\bibitem [{\citenamefont {Goryachev}\ \emph
  {et~al.}(2014{\natexlab{b}})\citenamefont {Goryachev}, \citenamefont {Farr},
  \citenamefont {Creedon}, \citenamefont {Fan}, \citenamefont {Kostylev},\ and\
  \citenamefont {Tobar}}]{YIG}%
  \BibitemOpen
  \bibfield  {author} {\bibinfo {author} {\bibfnamefont {M.}~\bibnamefont
  {Goryachev}}, \bibinfo {author} {\bibfnamefont {W.~G.}\ \bibnamefont {Farr}},
  \bibinfo {author} {\bibfnamefont {D.~L.}\ \bibnamefont {Creedon}}, \bibinfo
  {author} {\bibfnamefont {Y.}~\bibnamefont {Fan}}, \bibinfo {author}
  {\bibfnamefont {M.}~\bibnamefont {Kostylev}}, \ and\ \bibinfo {author}
  {\bibfnamefont {M.~E.}\ \bibnamefont {Tobar}},\ }\href {\doibase
  10.1103/PhysRevApplied.2.054002} {\bibfield  {journal} {\bibinfo  {journal}
  {Phys. Rev. Applied}\ }\textbf {\bibinfo {volume} {2}},\ \bibinfo {pages}
  {054002} (\bibinfo {year} {2014}{\natexlab{b}})}\BibitemShut {NoStop}%
\bibitem [{\citenamefont {Farr}\ \emph {et~al.}(2013)\citenamefont {Farr},
  \citenamefont {Creedon}, \citenamefont {Goryachev}, \citenamefont
  {Benmessai},\ and\ \citenamefont {Tobar}}]{parameters}%
  \BibitemOpen
  \bibfield  {author} {\bibinfo {author} {\bibfnamefont {W.~G.}\ \bibnamefont
  {Farr}}, \bibinfo {author} {\bibfnamefont {D.~L.}\ \bibnamefont {Creedon}},
  \bibinfo {author} {\bibfnamefont {M.}~\bibnamefont {Goryachev}}, \bibinfo
  {author} {\bibfnamefont {K.}~\bibnamefont {Benmessai}}, \ and\ \bibinfo
  {author} {\bibfnamefont {M.~E.}\ \bibnamefont {Tobar}},\ }\href {\doibase
  10.1103/PhysRevB.88.224426} {\bibfield  {journal} {\bibinfo  {journal} {Phys.
  Rev. B}\ }\textbf {\bibinfo {volume} {88}},\ \bibinfo {pages} {224426}
  (\bibinfo {year} {2013})}\BibitemShut {NoStop}%
\bibitem [{\citenamefont {Fano}(1961)}]{Fano}%
  \BibitemOpen
  \bibfield  {author} {\bibinfo {author} {\bibfnamefont {U.}~\bibnamefont
  {Fano}},\ }\href {\doibase 10.1103/PhysRev.124.1866} {\bibfield  {journal}
  {\bibinfo  {journal} {Phys. Rev.}\ }\textbf {\bibinfo {volume} {124}},\
  \bibinfo {pages} {1866} (\bibinfo {year} {1961})}\BibitemShut {NoStop}%
\bibitem [{\citenamefont {Bourhill}\ \emph {et~al.}(2013)\citenamefont
  {Bourhill}, \citenamefont {Benmessai}, \citenamefont {Goryachev},
  \citenamefont {Creedon}, \citenamefont {Farr},\ and\ \citenamefont
  {Tobar}}]{bourhill}%
  \BibitemOpen
  \bibfield  {author} {\bibinfo {author} {\bibfnamefont {J.}~\bibnamefont
  {Bourhill}}, \bibinfo {author} {\bibfnamefont {K.}~\bibnamefont {Benmessai}},
  \bibinfo {author} {\bibfnamefont {M.}~\bibnamefont {Goryachev}}, \bibinfo
  {author} {\bibfnamefont {D.~L.}\ \bibnamefont {Creedon}}, \bibinfo {author}
  {\bibfnamefont {W.}~\bibnamefont {Farr}}, \ and\ \bibinfo {author}
  {\bibfnamefont {M.~E.}\ \bibnamefont {Tobar}},\ }\href {\doibase
  10.1103/PhysRevB.88.235104} {\bibfield  {journal} {\bibinfo  {journal} {Phys.
  Rev. B}\ }\textbf {\bibinfo {volume} {88}},\ \bibinfo {pages} {235104}
  (\bibinfo {year} {2013})}\BibitemShut {NoStop}%
\end{thebibliography}

\end{document}